\documentclass[12pt]{article}
\usepackage{epsfig}

\usepackage{xcolor}

\usepackage{a4}
\usepackage{amsmath}

\usepackage{amssymb}
\usepackage{graphicx}
\usepackage{amsmath}
\usepackage{amsfonts} 

\newcommand{\la}{\lambda} 

\newcommand{\ka}{\kappa}
\newcommand{\f}{\phi}

\newcommand{\ta}{\theta}

\newcommand{\al}{\alpha}

\newcommand{\ee}{\end{equation}}
\newcommand{\eea}{\end{eqnarray}}
\newcommand{\be}{\begin{equation}}
\newcommand{\bea}{\begin{eqnarray}}

\newcommand{\pa}{\partial}
\newcommand{\Om}{\Omega}
\newcommand{\vep}{\varepsilon}

\newcommand{\om}{\omega}

\newcommand{\re}[1]{(\ref{#1})}
\newcommand{\R}{{\rm I \hspace{-0.52ex} R}}

\numberwithin{equation}{section}
\date{}

\begin{document}

\title{\bf Excited solutions in a Skyrme--Chern-Simons model in $2+1$ dimensions}

\author{ {\large Francisco Navarro-L\'erida}$^{\star}$
and 
{\large D. H. Tchrakian}$^{\dagger **}$
\\ 
\\
$^{\star}${\small  
Departamento de F\'isica Te\'orica and IPARCOS, Ciencias F\'isicas,
}\\
{\small Universidad Complutense de Madrid, E-28040 Madrid, Spain} 
\\  
$^{\dagger}${\small 
School of Theoretical Physics, Dublin Institute for Advanced Studies,}
\\
{\small  Burlington Road, Dublin 4, Ireland}
\\   
$^{**}${\small Department of Computer Science, Maynooth University, Maynooth, Ireland }
}

\maketitle

 \date{\today} 

\maketitle
\begin{abstract}
We study excited solutions in a Skyrme--Chern-Simons theory in $2+1$ dimensions. In particular, we emphasize the necessity of using a Lagrange multiplier method to obtain excited solutions, due to the appearance of a discontinuity when using a constraint compliant parametrization. These solutions are characterized by an integer number $p$, excited solutions corresponding to $p\neq 0$. The dependence of the global charges on the parameters is analyzed, showing non-standard behaviors. We also find that the presence of the Skyrme--Chern-Simons term does not alter significantly the pattern of energy levels, so $p=0$ solutions (fundamental solutions) have always the minimal energy.

\end{abstract}
\medskip
\medskip

\section{Introduction}
Chern-Simons (CS) densities, referred also as anomalies, have played an important role in many areas of gauge field
theory~\cite{Deser:1981wh,Deser:1982vy}. 
A salient feature of CS densities 
is that while they are not themselves invariant under gauge transformations,
their second-order equations of motion are gauge invariant.

A particular area of application, relevant to the subject of interest in the present work, has been in the construction of
solitons in the classical field theory of gauged  
Higgs~\cite{Paul:1986ix,Hong:1990yh,Jackiw:1990aw} and Skyrme~\cite{Schroers:1995he} fields in $2+1$ dimensions. 
Unusual effects were observed in $SO(2)$ gauged Skyrmions in $2+1$~\cite{Navarro-Lerida:2016omj,Navarro-Lerida:2018siw} and in
$4+1$~\cite{Navarro-Lerida:2020hph} dimensions. These  unusual
effects include the appearance of {\it both positive and negative slopes} in the plots of $(E,J)$ and $(E,Q_e)$, where
$E$ is the mass/energy, $J$ the angular momentum, and $Q_e$ the electric charge of the $SO(2)$ gauged Skyrmion. In addition it
was seen that the presence
of the CS term in these systems led to a change in the topological charge, namely the ``baryon number'', of the Skyrmions.
We will return to these effects when analysing the properties of the generalisation of CS anomalies below.

Chern-Simons densities result from Chern-Pontryagin (CP) densities, the latter being both gauge-invariant and a total-divergence.
The CS densities are defined only in odd dimensions since they result
from a one-step descent of CP densities, which themselves are defined only in even dimensions. This aspect of CS densities
excludes the application of anomalies to field theories in even spacetime dimensions. It is to tackle this obstacle, that 
Skyrme--Chern-Simons (SCS) densities were proposed in Ref.~\cite{Tchrakian:2015pka}. The SCS are analogues of the CS, whose
definition follows from a one-step descent of
Skyrme--Chern-Pontryagin~\cite{Tchrakian:2024bqy} (SCP) densities, the latter being both gauge-invariant and total-divergence,
and like CS anomalies, the SCS densities are not themselves gauge invariant but result in gauge invariant equations of motion.
But in contrast to the CP densities which are defined only in even dimensions,
the SCP densities~\cite{Tchrakian:2024bqy} are defined in all dimensions -- even and odd.
Hence the SCS densities resulting from the one-step
descent of the latter, are also defined in all odd and even dimensions.

To date we have studied the simplest SCS density in $2+1$ dimensions in~\cite{Navarro-Lerida:2023fsr} (in the absence of the
usual CS term) as a prototype, to test whether the unusual properties resulting from the effect of the CS term on the
$SO(2)$ gauged $O(3)$ Skyrmion observed in $2+1$ dimensions~\cite{Navarro-Lerida:2016omj,Navarro-Lerida:2018siw} would appear also with a SCS term. It was
observed that these effects were largely similar in both cases. The next example, in~$3+1$ dimensions, is under active consideration.
The purpose of the present work is to study details of the earlier work~\cite{Navarro-Lerida:2023fsr} in $2+1$ dimensions, to
resolve a remaining question with a view to proceeding to $3+1$ dimensions.

From the viewpoint of applications to physics, clearly the most important case is that of $3+1$ dimensions.
The problem of constructing anomaly densities for gauged Skyrme fields in $3+1$ dimensions has been active since a long time, and it may
be in order to comment on it briefly here. The approach was based on gauging the  Wess-Zumino (WZ) action as formulated in
Ref.~\cite{Witten:1983tw} and in a vast literature thereafter (see,
$e.g.,$ \cite{Brihaye:1984hp,Figueroa-OFarrill:1994vwl}). Subsequently, the work of Ref.~\cite{Witten:1983tw} was specialised to gauge
group $SO(2)$ and an explicit formula for the anomaly in that case was given in~\cite{Callan:1983nx}, to which we refer here as the
Callan-Witten (CW) anomaly. More recently a detailed analysis of the model consisting of the $SO(2)$ gauged Skyrme action in $3+1$
dimensions in the presence of the CW anomaly, was carried out
in Ref.~\cite{Navarro-Lerida:2023hbv}, where Chern-Simons like effects like those seen in \cite{Navarro-Lerida:2016omj,Navarro-Lerida:2018siw}
on the gauged Skyrmion in $2+1$ dimensions were observed. The corresponding
analysis employing the SCS density instead of the CW anomaly in $3+1$, has not been carried out to date. Indeed, it is with this in mind
that we revisit the SCS action in $2+1$ dimensions, whose effects on the Abelian gauged planar Skyrmion were studied
in~\cite{Navarro-Lerida:2023fsr} previously, where certain dynamical effects deserving further study were encountered.

In contrast to the gauged WZ actions~\cite{Witten:1983tw,Brihaye:1984hp,Figueroa-OFarrill:1994vwl,Callan:1983nx} which are strictly in $3+1$ dimensions,
the definition of the SCS density arises from a prescription which applies to all dimensions. A noteworthy difference between the gauged Wess-Zumino-Witten (WZW)
actions in $3+1$ dimensions and its SCS counterpart is that in the latter the WZ term itself appears.
This is because in that case the full Skyrme scalar is that of the $O(6)$ sigma model, rather than the $O(4)$
scalar describing the Skyrmion. This particular property of SCS actions will be the centre of the present note.

To test the analogy between the effects of the CS and SCS actions, we have in Ref.~\cite{Navarro-Lerida:2023fsr}
studied the system consisting of the action that supports
the $SO(2)$ gauged Skyrmion in $2+1$ dimensions, in the presence of the SCS action (and in the absence of the CS term).
It was verified that the exotic features observed for the the plots $(E,J)$ and $(E,Q_e)$ in the presence
of the CS term, were observed qualitatively also in the presence of the SCS term.

While the ultimate aim would be the study of the effect of the SCS term on the gauged Skyrmion in $3+1$ dimensions, our initial study was
carried out for the $2+1$ system studied in Ref.~\cite{Navarro-Lerida:2023fsr}, as a prototype for the $3+1$ dimensional case. Technically,
the system describing the Abelian (Maxwell) field on $\R^2$ in the static limit can be subjected to azimuthal symmetry, resulting in a
one dimensional action density. The resulting field equations then, can be described by ODE's in the radial coordinate $r$. By contrast,
the system describing the Abelian~\footnote{The SCS density in $3+1$ dimensions is encoded with either an $SO(2)\times SO(2)$ or an
$SO(2)\times SO(3)$ connection~\cite{Tchrakian:2024bqy}.} field on $\R^3$ in the static limit when subjected to azimuthal symmetry results
in a two dimensional reduced action described by PDE's in the radial and polar coordinates $(r,\ta)$.

Having settled to the study of the prototype in $2+1$ dimensions before proceeding to the $3+1$ dimensional case, we return to details of
the mechanism encountered in Ref.~\cite{Navarro-Lerida:2023fsr}. The system there was described by the fields $(A_\mu,\f^a\ ,\ a=1,2,3)$, 
the $SO(2)$ gauge connection and the $O(3)$ Skyrme scalar supporting the gauged planar Skyrmion, and 
$(A_\mu,B_\mu,\ta^{\tilde a}\ ,\ \tilde a=1,2,\dots,5)$ which encode the SCS action, where $B_\mu$ was described as the auxiliary Abelian field
and $\ta^{\tilde a}$ as the auxiliary $O(5)$ Skyrme scalar. The results in Ref.~\cite{Navarro-Lerida:2023fsr} were restricted to that of
a contracted model, in which the $O(5)$ Skyrme scalar
\be
\ta^{\tilde a}=(\ta^\al,\ta^A,\ta^5)\ ,\quad\al=1,2\ \ {\rm and}\ \ A=3,4\ ,
\ee
is contracted to an $O(3)$ field via the contraction
\be
\ta^{\tilde a}\,\to\,\ta^a\stackrel{\rm def.}=(\ta^\al,\ta^5)\,,\quad a=\al,5 \ ,\label{contr}
\ee
by setting $\ta^A=0$.

The choice \re{contr} is quite natural since the ungauged sigma model scalar $\ta^a\in{O(3)}$ can support
topologically stable solutions on $\R^2$~\cite{Manton:2004tk}, while
an $O(5)$ sigma model encoded by the scalar $\ta^{\tilde a}$ does not. This argument is independent of whether the Skyrme scalar is
gauged or not, so it applies equally well to the $SO(2)\times SO(2)$ gauged system defining the SCS density. Hence, it would be expected
that the field configurations resulting from a system described by $\ta^{\tilde a}$ will have higher energy and be unstable against
decay to the solutions of the system described by $\ta^a$.

Retaining the full $O(5)$ Skyrme scalar $\ta^{\tilde a}$ is akin to the situation of the Weinberg-Salam model on $\R^3$,
where the $SU(2)$ gauged complex doublet Higgs scalar also cannot support finite energy topologically stable solutions~\cite{Manton:2004tk}. Rather, the system supports ``sphaleron'' solutions~\cite{Manton:1983nd,Klinkhamer:1984di}, which are not topologically
stable because of the multiplet structure of the Higgs scalar, which is encoded with {\bf one} extraneous parameter that results in the
instability. Here by contrast the $O(5)$ Skyrme scalar is encoded with {\bf two} such parameters $(\ta^3,\ta^4)$, which renders
geometric treatment as in Ref.~\cite{Manton:1983nd} difficult. We have thus opted to analyse the ODE's of the full system, namely the
$SO(2)\times SO(2)$ gauged $O(5)$ Skyrme scalar.

Our method of constructing the solutions to the full gauged $O(5)$ system is anchored on the solutions in the
gauge decoupling limit of the model. Hence the model employed here differs from the one in Ref.~\cite{Navarro-Lerida:2023fsr}, in that
in addition to the quadratic Skyrme kinetic term, there appears also the quartic Skyrme kinetic term necessary for satisfying
Derrick scaling in the gauge decoupled model.

We refer to the solutions of the model described by the scalar $\ta^a$ as the $fundamental$ solution, while those described by $\ta^{\tilde a}$
as $excited$ solutions. Computationally, such solutions are first constructed for the ungauged model, which is then gauged with the $SO(2)\times SO(2)$
gauge field.

In addition to starting from the ungauged model, we also truncate the full system further by eliminating the
$O(3)$ Skyrme scalar $\f^a$ which would support the soliton itself. Thus, we restrict our investigation to the full SCS density together with both
kinetic terms of the gauge fields and the Skyrme kinetic terms, the latter 
including both the quadratic and quartic members.

The paper is organized as follows: We introduce the model with its Lagrangian, static radial Ansatz and appropriate boundary conditions in Section 2, the computation of the global charges is addressed in Section 3, the numerical results are presented in Section 4 and the conclusions are discussed in Section 5.

\section{The model}
In Ref.~\cite{Navarro-Lerida:2023fsr} we introduced an $SO(2)$ gauged $O(3)$ sigma model, supplemented with an $O(5)$ Skyrme scalar gauged also with an “auxiliary” Abelian $SO(2)$ field, including in the dynamics a SCS term. There the model was contracted further so that the $O(5)$ Skyrme scalar became an effective $O(3)$ scalar. Here in order to isolate the effect of the SCS term and to analyze fully excited solutions, we consider an $SO(2) \times SO(2)$ gauged $O(5)$ Skyrme model in $2+1$ dimensions endowed with the corresponding SCS term. The potential employed is a ``pion mass'' potential.   

The Lagrangian of the system we study here is
\be
\label{L}
{\cal L}= -\frac14\, F_{\mu\nu}^2 -\frac14\, G_{\mu\nu}^2 +\frac12\,|D_\mu\ta^{\bar a}|^2 -\frac{1}{8} |D_{[\mu} \ta^{\bar a} D_{\nu]} \ta^{\bar b} |^2 - \mu(1-\ta^5 ) + \ka\,\Om_{\rm SCS} \,, 
\ee
where the $SO(2) \times SO(2)$ gauge field is defined by
\be
\label{gauge_field}
F_{\mu \nu} = \partial_\mu A_\nu - \partial_\nu A_\mu \, , \, \, \, G_{\mu \nu} = \partial_\mu B_\nu - \partial_\nu B_\mu  \,, 
\ee
and the $O(5)$ scalar is
\be
\label{scalar}
(\ta^{\bar a})=\left(\begin{array}{l}
\ta^1\\
\ta^2\\
\ta^3\\
\ta^4\\
\ta^5
\end{array}\right)
\equiv
\left(\begin{array}{l}
\ta^{\al}\\
\ta^{A}\\
\ta^{5}
\end{array}\right) \, ,
\ee
with ${\bar a}=1,\dots,5$ ($\al=1,2$, $A=3,4$), satisfying the constraint
\be
\label{contraint}
|\ta^{\bar a}|^2=1 \, .
\ee

The gauge covariant derivatives of the scalar are defined by
\bea
D_\mu\ta^\al&=&\pa_\mu\ta^\al+A_\mu\,(\vep\ta)^\al \,, \ \ \ \alpha=1,2 \,,
\label{Dtaal}
\\
D_\mu\ta^A&=&\pa_\mu\ta^A+B_\mu\,(\vep\ta)^A \,, \ \ \ A=3,4 \,,
\label{DtaA}
\\
D_\mu\ta^5&=&\pa_\mu\ta^5\,,\label{Dta5}
\eea
where 
$(\vep\ta)^1 =\ta^2, (\vep\ta)^2 =-\ta^1, (\vep\ta)^3 =\ta^4, (\vep\ta)^4 =-\ta^3$.

The $SCS$ term, $\Om_{\rm SCS}$, can be written as
\be
\label{SCS}
\Om_{\rm SCS}=\om+\Om\,,
\ee
as given in Ref.~\cite{Tchrakian:2021xzy}. The second term, $\Om$, can be given explicitly 
\bea
\Om&=&3!\,\vep^{\mu\nu\la}\ \ta^5\bigg\{\frac13\,(\ta^5)^2\nonumber
(A_\la G_{\mu\nu}+B_\la F_{\mu\nu})
-A_\mu B_\nu\ \pa_\la(|\ta^\al|^2-|\ta^A|^2)\nonumber\\
&&
-2\left[A_\la\ (\vep\pa_\mu\ta)^A\,\pa_\nu\ta^A+B_\la\ (\vep\pa_\mu\ta)^\al\,\pa_\nu\ta^\al\right]\bigg\}\, ,\label{SCSAB}
\eea
while the first one, namely, the Wess-Zumino term, $\omega$, can only be expressed in  {\it constraint compliant} parametrization.

\subsection{Constraint compliant parametrization}
The constraint compliant parametrization of the
Skyrme scalar is expressed by
\be
\label{cct}
(\ta^{\bar a})=\left(\begin{array}{l}
\ta^{\al}\\
\ta^{A}\\
\ta^{5}
\end{array}\right)
=
\left(\begin{array}{l}
 \sin f(x_\mu)\sin g(x_\mu)\ n^{\al}\\
\sin f(x_\mu)\cos g(x_\mu)\ m^{A}\\
\cos f(x_\mu)
\end{array}\right) 
\,,
\ee
where
$n^\al$ and $m^A$ are parametrized as
\be
\label{nal}
(n^{\al})=\left(\begin{array}{l}
\cos\psi(x_\mu)\\
\sin\psi(x_\mu)\\
\end{array}\right)\, ,\quad
(m^{A})=\left(\begin{array}{l}
\cos\chi(x_\mu)\\
\sin\chi(x_\mu)\\
\end{array}\right)\,. 
\ee

In this parametrization, the full SCS density \re{SCS} is expressed as
\bea
\label{SCSgaugecomp}
\Om_{\rm SCS}&=&3!\vep^{\mu\nu\la}\bigg\{
-2\,\left(\cos f-\frac13\cos^3f
\right)\,(\pa_\la\sin^2g)\,\pa_\mu\psi\,\pa_\nu\chi\nonumber\\
&&\quad\quad\quad+\frac13\,\cos^3f\,(A_\la G_{\mu\nu}+B_\la F_{\mu\nu})\nonumber\\
&&\quad\quad\quad-A_\mu B_\nu\,\cos f\left[\pa_\la(\sin^2f\sin^2g)-\pa_\la(\sin^2f\cos^2g)\right]\nonumber\\
&&\quad\quad\quad 
+2\cos f\left[A_\la\,\pa_\mu(\sin^2f\cos^2g)\ \pa_\nu\chi+B_\la\,\pa_\mu(\sin^2f\sin^2g)\ \pa_\nu\psi\right]\bigg\}\,,
\eea
the first line of which is $\om$ in \re{SCS}, which obviously vanishes in the static limit.

\subsection{Field equations and static radial Ansatz}
The general field equation can be obtained by taking variations of \re{L} with respect to $A_\mu, B_\mu, f, g, \psi$, and $\chi$ (with $\Om_{\rm SCS}$ given in constraint compliant parametrization \re{SCSgaugecomp}). The presence of the Wess-Zumino term in \re{SCS} prevents us from using a Lagrange multiplier method and avoid the use of the constraint compliant parametrization. However, for static radial solutions, the Wess-Zumino term $\omega$ does not enter the field equations and one can formally obtain those equations using a Lagrange multiplier method over a Lagrangian where $\Om_{\rm SCS}$ is substituted effectively by $\Om$ \re{SCSAB}. That approach has the advantage to skip the computation of the function $g$ directly, which has been found to be numerically difficult.

The static radial Ansatz we will employ for the gauge potential can be written as
\bea
A_0=c(r) \, , & \displaystyle A_i=\left(\frac{a(r)-n}{r}\right)(\vep\hat x)_i \, , \ \ \ i=1,2 \,,\label{MaxaxA} \\
B_0=d(r) \, , & \displaystyle B_i=\left(\frac{b(r)-m}{r}\right)(\vep\hat x)_i \, , \ \ \ i=1,2 \,,\label{MaxaxB}
\eea
with $\hat x_i \equiv x_i/r$, $i=1,2$ and $(\vep\hat x)_1=\hat x_2 \ , \ (\vep\hat x)_2=-\hat x_1$, in polar coordinates $(r,\varphi)$ (with $x_1=r\cos\varphi, x_2=r\sin\varphi$).

For the $O(5)$ scalar, the azimuthal symmetry is expressed by restricting the functions $f, g, \psi$, and $\chi$ to
\be
\label{scalar_r}
f=f(r) \, , g=g(r) \, , \psi = n\varphi \, , \chi = m\varphi \, ,  
\ee 
$n$ and $m$ being non-vanishing integer numbers.

As mentioned above, for this type of solutions it is useful to use a non-constraint compliant parametrization by defining the functions
\bea
&&R(r)=\sin(f(r)) \sin(g(r)) \, , \label{function_R} \\
&&S(r)=\sin(f(r)) \cos(g(r)) \, , \label{function_S} \\
&&T(r)=\cos(f(r)) \, , \label{function_T}
\eea
satisfying
\be
R(r)^2+S(r)^2+T(r)^2=1 \, . \label{constraint_RST}
\ee

The resulting system of ODEs possesses simplified solutions for the cases $g(r)=0$ (equivalently, $R(r)=0$) or $g(r)=\pi/2$ (equivalently, $S(r)=0$), for which the $O(5)$ scalar reduces to an effective $O(3)$ scalar.

\subsection{Boundary conditions and expansions}

We are interested in finite-energy configurations with azimuthal symmetry. More precisely, we want to obtain excited solutions for the full $O(5)$ scalar field, so $g(r)$ will not be constant, in general. The analysis of the field equations reveals that it is necessary that $|n|\neq |m|$, with $|n|, |m| \ge 2$. Due to the symmetry $(n,m) \leftrightarrow (-n,-m)$ and under the interchange $(n,m) \leftrightarrow (m,n)$, without loss of generality, in order to obtain excited solutions we will assume
\be
n > m \ge 2 \, . \label{n_m_gt_2}
\ee

The boundary conditions at the origin read
\bea
a(0) = n \, , &c'(0) = 0 \, , \label{BCs_A_0} \\
b(0) = m \, , &d'(0) = 0 \, , \label{BCs_B_0} \\
R(0) = 0\, , &S(0) = 0\, , &T(0) = -(-1)^p \, , \label{BCs_RST_0}
\eea
with $p=0,1,2,\dots$ (Remark: Negative integer values of $p$ result in equivalent solutions to those with positive values.) The corresponding boundary conditions in constraint compliant parametrization read
\bea
f(0)=\frac{1+(-1)^p}{2}\pi \, , &g(0)=p \pi \, . \label{BCs_fg_0}
\eea

This set of boundary conditions is a consequence of the expansions at the origin of the functions that ensure regularity of the solutions and finite energy. These expansions read
\bea
a(r) &=& n + a_2 r^2 + O(r^4) \, , \label{exp_or_a} \\
c(r) &=& c_0 + O(r^2) \, , \label{exp_or_c} \\
b(r) &=& m + b_2 r^2 + O(r^4) \, , \label{exp_or_b} \\
d(r) &=& d_0 + O(r^2) \, , \label{exp_or_d} \\
R(r) &=& R_n r^n + O(r^{n+2}) \, , \label{exp_or_R} \\
S(r) &=& S_m r^m + O(r^{m+2}) \, , \label{exp_or_S} \\
T(r) &=& -(-1)^p + \frac{1}{2} (-1)^p S_m^2 r^{2 m} + O(r^{2m+2}) \, , \label{exp_or_T}
\eea
whereas for the functions $f(r)$ and $g(r)$ they are
\bea
f(r) &=& \frac{1+(-1)^p}{2}\pi + f_m r^m + O(r^{m+2}) \, , \label{exp_or_f} \\
g(r) &=& p \pi + g_{n-m} r^{n-m} + O(r^{n-m+2}) \, , \label{exp_or_g}
\eea
with
\bea
R_n = -f_m g_{n-m} \, , & S_m = -f_m \, . \label{rels_or}
\eea

The boundary conditions at infinity are
\bea
\lim_{r \to \infty} a(r)=a_\infty \, ,  &\displaystyle \lim_{r \to \infty} c(r) = c_\infty \, ,  \label{BCs_A_infty} \\
\lim_{r \to \infty} b(r)=b_\infty \, ,  &\displaystyle \lim_{r \to \infty} d(r) = d_\infty \, ,  \label{BCs_B_infty} \\
\lim_{r \to \infty} R(r)=0 \, ,  &\displaystyle \lim_{r \to \infty} R(r) = 0 \, , & \lim_{r \to \infty} T(r)=1 \, .\label{BCs_RST_infty}
\eea

For $f(r)$ and $g(r)$ the boundary conditions are a bit more involved since they depend on the relative magnitude of $|c_\infty|$ and $|d_\infty|$. There are three possibilities:
\bea
\lim_{r \to \infty} f(r)=0 \, ,  &\displaystyle \lim_{r \to \infty} g(r) = 0 \, , \quad  {\rm for} \ |c_\infty| < |d_\infty| \, , \label{BCs_fg_infty_1} \\
\lim_{r \to \infty} f(r)=0 \, ,  &\displaystyle \lim_{r \to \infty} g(r) = \frac{\pi}{2} \, , \quad  {\rm for} \ |c_\infty| > |d_\infty| \, , \label{BCs_fg_infty_2}\\
\lim_{r \to \infty} f(r)=0 \, ,  &\displaystyle \lim_{r \to \infty} g(r) = g_\infty \, , \quad  {\rm for} \ |c_\infty| = |d_\infty| \, , \label{BCs_fg_infty_3}
\eea
with $g_\infty \neq 0$ or $\pi/2$. Owing to that the asymptotic expansions of excited solutions need to be considered separately in these three regimens:

{\bf Case $|c_\infty| < |d_\infty|$}
\bea
a(r)&=&a_\infty + c_1 \sqrt{r} e^{-8\kappa} + {\hat c}_1 e^{-2\sqrt{\mu-d_\infty^2}r} + \dots \, , \label{exp_inf_a_1} \\
c(r)&=&c_\infty + c_2 \frac{1}{\sqrt{r}} e^{-8\kappa} + {\hat c}_2\frac{1}{r^2} e^{-2\sqrt{\mu-d_\infty^2}r} + \dots \, , \label{exp_inf_c_1} \\
b(r)&=&b_\infty + c_2 \sqrt{r} e^{-8\kappa} + {\hat c}_3 \frac{1}{r}e^{-2\sqrt{\mu-d_\infty^2}r} + \dots \, , \label{exp_inf_b_1} \\
d(r)&=&d_\infty + c_1 \frac{1}{\sqrt{r}} e^{-8\kappa} + {\hat c}_4\frac{1}{r} e^{-2\sqrt{\mu-d_\infty^2}r} + \dots \, , \label{exp_inf_d_1} \\
R(r) &=& c_3 c_4 \frac{1}{\sqrt{r}}  e^{-\sqrt{\mu-c_\infty^2}r} + \dots \, , \label{exp_inf_R_1} \\
S(r) &=& c_3 \frac{1}{\sqrt{r}}  e^{-\sqrt{\mu-d_\infty^2}r} + \dots \, , \label{exp_inf_S_1}\\
T(r) &=& 1-\frac{c_3^2}{2} \frac{1}{r}  e^{-2\sqrt{\mu-d_\infty^2}r} + \dots \, , \label{exp_inf_T_1}
\eea
whereas for $f(r)$ and $g(r)$
\bea
f(r) &=& c_3 \frac{1}{\sqrt{r}}  e^{-\sqrt{\mu-d_\infty^2}r} + \dots \, , \label{exp_inf_f_1} \\
g(r) &=& c_4  e^{-\left(\sqrt{\mu-c_\infty^2}-\sqrt{\mu-d_\infty^2}\right)r} + \dots \, . \label{exp_inf_g_1}
\eea

{\bf Case $|c_\infty| > |d_\infty|$}
\bea
a(r)&=&a_\infty + c_1 \sqrt{r} e^{-8\kappa} + {\hat c}_1 \frac{1}{r} e^{-2\sqrt{\mu-c_\infty^2}r} + \dots \, , \label{exp_inf_a_2} \\
c(r)&=&c_\infty + c_2 \frac{1}{\sqrt{r}} e^{-8\kappa} + {\hat c}_2\frac{1}{r} e^{-2\sqrt{\mu-c_\infty^2}r} + \dots \, , \label{exp_inf_c_2} \\
b(r)&=&b_\infty + c_2 \sqrt{r} e^{-8\kappa} + {\hat c}_3 e^{-2\sqrt{\mu-c_\infty^2}r} + \dots \, , \label{exp_inf_b_2} \\
d(r)&=&d_\infty + c_1 \frac{1}{\sqrt{r}} e^{-8\kappa} + {\hat c}_4\frac{1}{r^2} e^{-2\sqrt{\mu-c_\infty^2}r} + \dots \, , \label{exp_inf_d_2} \\
R(r) &=& c_3 \frac{1}{\sqrt{r}}  e^{-\sqrt{\mu-c_\infty^2}r} + \dots \, , \label{exp_inf_R_2} \\
S(r) &=& -c_3c_4 \frac{1}{\sqrt{r}}  e^{-\sqrt{\mu-d_\infty^2}r} + \dots \, , \label{exp_inf_S_2}\\
T(r) &=& 1-\frac{c_3^2}{2} \frac{1}{r}  e^{-2\sqrt{\mu-c_\infty^2}r} + \dots \, , \label{exp_inf_T_2}
\eea
whereas for $f(r)$ and $g(r)$
\bea
f(r) &=& c_3 \frac{1}{\sqrt{r}}  e^{-\sqrt{\mu-c_\infty^2}r} + \dots \, , \label{exp_inf_f_2} \\
g(r) &=&\frac{\pi}{2}+ c_4  e^{-\left(\sqrt{\mu-d_\infty^2}-\sqrt{\mu-c_\infty^2}\right)r} + \dots \, . \label{exp_inf_g_2}
\eea

{\bf Case $|c_\infty| = |d_\infty|$}
\bea
a(r)&=&a_\infty + c_1 \sqrt{r} e^{-8\kappa} + {\hat c}_1 e^{-2\sqrt{\mu-c_\infty^2}r} + \dots \, , \label{exp_inf_a_3} \\
c(r)&=&c_\infty + c_2 \frac{1}{\sqrt{r}} e^{-8\kappa} + {\hat c}_2\frac{1}{r} e^{-2\sqrt{\mu-c_\infty^2}r} + \dots \, , \label{exp_inf_c_3} \\
b(r)&=&b_\infty + c_2 \sqrt{r} e^{-8\kappa} + {\hat c}_3 e^{-2\sqrt{\mu-c_\infty^2}r} + \dots \, , \label{exp_inf_b_3} \\
d(r)&=&c_\infty + c_1 \frac{1}{\sqrt{r}} e^{-8\kappa} + {\hat c}_4\frac{1}{r} e^{-2\sqrt{\mu-c_\infty^2}r} + \dots \, , \label{exp_inf_d_3} \\
R(r) &=& c_3 \sin(g_\infty) \frac{1}{\sqrt{r}}  e^{-\sqrt{\mu-c_\infty^2}r} + \dots \, , \label{exp_inf_R_3} \\
S(r) &=& c_3\cos(g_\infty) \frac{1}{\sqrt{r}}  e^{-\sqrt{\mu-c_\infty^2}r} + \dots \, , \label{exp_inf_S_3}\\
T(r) &=& 1-\frac{c_3^2}{2} \frac{1}{r}  e^{-2\sqrt{\mu-c_\infty^2}r} + \dots \, , \label{exp_inf_T_3}
\eea
whereas for $f(r)$ and $g(r)$
\bea
f(r) &=& c_3 \frac{1}{\sqrt{r}}  e^{-\sqrt{\mu-c_\infty^2}r} + \dots \, , \label{exp_inf_f_3} \\
g(r) &=& g_\infty + \frac{ (a_\infty^2-b_\infty^2)\sin(2 g_\infty)}{4\sqrt{\mu -c_\infty^2}} \frac{1}{r} + \dots \, . \label{exp_inf_g_3}
\eea
(Remark: Given $\mu,  \kappa, m, n$, and $p$, the only free parameters are $c_\infty$ and $d_\infty$, the rest of parameters being numerically determined.)

\section{Global charges}

\subsection{Energy and angular momentum}
The stress-energy tensor associated to the Lagrangian \re{L} can be computed using the GR prescription, namely,
\be
T_{\mu \nu} =  -2 \frac{\partial {\cal L}}{\partial \eta_{\mu \nu}} + \eta_{\mu \nu} {\cal L} \, ,\label{SE_tensor}
\ee
with $(\eta_{\mu\nu})=diag(1,-1,-1)$, the Minkowski metric in Lorentzian coordinates $(x_0,x_1,x_2)$. The explicit form of the stress-energy tensor reads
\bea
T_{\mu \nu} &=& -\left(F_{\mu\rho} {F^\rho}_\nu + \frac{1}{4} \eta_{\mu\nu} F_{\rho\sigma}F^{\rho \sigma}\right)  -\left(G_{\mu\rho} {G^\rho}_\nu + \frac{1}{4} \eta_{\mu\nu} G_{\rho\sigma}G^{\rho \sigma}\right) \nonumber \\
&&-\left(D_\mu \theta^{\bar a} D_\nu \theta^{\bar a} -\frac{1}{2}\eta_{\mu\nu} D_\rho \theta^{\bar a} D^\rho \theta^{\bar a}\right) \nonumber \\
&&-\frac{1}{2}\left[ \left( D_{[\mu} \theta^{\bar a} D_{\rho]}\theta^{\bar b} \right)   \left( D_{[\sigma} \theta^{\bar a} D_{\nu]}\theta^{\bar b} \right)\eta^{\rho \sigma}  + \frac{1}{4} \eta_{\mu \nu} \left( D_{[\rho} \theta^{\bar a} D_{\sigma]}\theta^{\bar b} \right)  \left( D^{[\rho} \theta^{\bar a} D^{\sigma]}\theta^{\bar b} \right) \right] \nonumber \\
&&-\eta_{\mu\nu} \, \mu(1-\theta^5) \, . \label{T_munu}
\eea
Notice that the SCS density does not enter the stress-energy tensor.

From \re{T_munu} we can compute the energy density, $\cal E$, and the angular momentum density, $\cal J$, defined, respectively, by
\bea
\cal E &\equiv& - T_{00} \, , \label{E_dens}\\
\cal J &\equiv& x_1 T_{02} -x_2 T_{01} \, . \label{J_dens}
\eea

For the radial solutions we are considering here \re{MaxaxA}-\re{function_T}, these densities read
\bea
\cal E &=& \frac{1}{2} \left[ (c')^2 + \frac{1}{r^2}(a')^2\right] +  \frac{1}{2} \left[ (d')^2 + \frac{1}{r^2}(b')^2\right] + \nonumber \\
&& \frac{1}{2} \left[ \left(c^2+\frac{1}{r^2}a^2\right)R^2 + \left(d^2+\frac{1}{r^2}b^2\right)S^2 + (R')^2 + (S')^2 + (T')^2 \right] +\nonumber \\
&& \frac{1}{2} \left[ \left(\left(c^2+\frac{1}{r^2}a^2\right)R^2 + \left(d^2+\frac{1}{r^2}b^2\right)S^2\right)( (R')^2 + (S')^2 + (T')^2) + \nonumber \right. \\
  && \left. \frac{1}{r^2}(a d- b c)^2 R^2 S^2 \right] + \mu (1-T) \, , \label{E_dens_ansatz}
\eea
\bea
\cal J &=& a' c' + b' d' + ac R^2 + bd S^2 + (ac R^2 +bd S^2)((R')^2 + (S')^2 + (T')^2) \, , \label{J_dens_ansatz}
\eea
where in constraint compliant parametrization they become
\bea
\cal E &=& \frac{1}{2} \left[ (c')^2 + \frac{1}{r^2}(a')^2\right] +  \frac{1}{2} \left[ (d')^2 + \frac{1}{r^2}(b')^2\right] + \nonumber \\
&&  \frac{1}{2} \left[ (f')^2 +\sin^2 f \, (g')^2 +\sin^2 f \left(\left(c^2+\frac{1}{r^2} a^2\right) \sin^2 g + \left(d^2+\frac{1}{r^2} b^2\right) \cos^2 g \right) \right] + \nonumber \\
&&  \frac{1}{2} \sin^2 f \left[\frac{1}{r^2} \sin^2 f \sin^2 g \cos^2 g (ad-bc)^2 + \right. \nonumber \\
  && \left.   \left(\left(c^2+\frac{1}{r^2} a^2\right) \sin^2 g + \left(d^2+\frac{1}{r^2} b^2\right) \cos^2 g \right) \left( (f')^2 + \sin^2 f \, (g')^2 \right) \right] +\nonumber \\
&&  \mu (1-\cos f) \, , \label{E_dens_ansatz_ccp}
\eea
\bea
\cal J &=& a' c' + b' d' + \left[ 1 + (f')^2 +\sin^2 f \, (g')^2 \right] \sin^2 f \, (ac\sin^2 g +bd \cos^2 g) \, . \label{J_dens_ansatz_ccp}
\eea 

This last expression for $\cal J$ can be written in a more useful form using the field equations in constraint compliant parametrization, namely,
\bea
\cal J &=& \frac{1}{r} \frac{d}{dr} \left[ 8 \kappa a b \cos^3 f + r ac' + r b d' \right] \, .  \label{J_dens_ansatz_ccp_2}
\eea

The energy, $E$, and the angular momentum, $J$, of the solutions can be readily obtained from the corresponding densities
\be
E = \int_{\R^2} {\cal E} d^2x = 2\pi \int_0^\infty r {\cal E} dr \, , \label{energy}
\ee
\be
J = \int_{\R^2} {\cal J} d^2x = 2\pi \int_0^\infty r {\cal J} dr = 16 \pi \kappa \left[ a_\infty b_\infty + (-1)^p n m \right] \, , \label{ang_mom} 
\ee
where the last term in \re{ang_mom} results from \re{BCs_A_0}-\re{BCs_fg_0}, \re{BCs_A_infty}-\re{BCs_fg_infty_3}, and \re{J_dens_ansatz_ccp_2}.

\subsection{Electric charges}
In order to define the electric charges we will follow a Noether approach. Using the constraint complaint parametrization \re{cct}-\re{nal}, it is easy to see that under a gauge transformation
\bea
A_\mu &\to& A_\mu +\partial_\mu \Lambda_1 \, , \\
\psi &\to& \psi + \Lambda_1\, , \\
B_\mu &\to& B_\mu + \partial_\mu \Lambda_2 \, , \\
\chi &\to & \chi + \Lambda_2 \, ,
\eea
the Lagrangian remains unchanged and so does the action, so the following two currents are conserved
\bea
j_{(1)}^\mu \equiv \frac{\partial \cal L}{\partial \psi_\mu} \, , \quad j_{(2)}^\mu \equiv \frac{\partial \cal L}{\partial \chi_\mu} \, . \label{currents}
\eea
We define the charge densities associated to $A_\mu$ and $B_\mu$ by
\bea
\rho_e \equiv -j_{(1)}^0 \, , \quad  \rho_g \equiv -j_{(2)}^0 \, . \label{charge_dens} 
\eea

For the radial solutions we are considering here \re{MaxaxA}-\re{function_T} and using the field equations, these densities can be written as
\bea
\rho_e &=& \frac{1}{r} \frac{d}{dr} \left[ r c' + 8\kappa (b-m\sin^2 g) \cos^3f\right] \, , \label{rho_e}\\
\rho_g &=& \frac{1}{r} \frac{d}{dr} \left[ r d' + 8\kappa (a-n\cos^2 g) \cos^3f\right] \, . \label{rho_g}
\eea

The electric charges result to be
\bea
&&Q_e =  \int_{\R^2} \rho_e d^2x = 2\pi \int_0^\infty r \rho_e dr =16\pi\kappa \left[ b_\infty + m [(-1)^p -\sin^ 2g(\infty)]\right] \, , \label{Q_e}\\
&&Q_g =  \int_{\R^2} \rho_g d^2x = 2\pi \int_0^\infty r \rho_g dr  =16\pi\kappa \left[ a_\infty - n \cos^ 2g(\infty)\right] \, , \label{Q_g}
\eea
where we have used \re{BCs_A_0}-\re{BCs_fg_0}. From \re{BCs_fg_infty_1}-\re{BCs_fg_infty_3} it is clear that these charges are discontinuous at $|c_\infty| =|d_\infty|$. As we will see in the following section, the other quantities are continuous at  $|c_\infty| =|d_\infty|$. In fact, this anomaly is easily solved by considering the total electric charge, $Q_t$, which is continuous at $|c_\infty| =|d_\infty|$
\be
Q_t = n Q_e + m Q_g = 16\pi\kappa \left[ n b_\infty +m a_\infty + n m ((-1)^p -1)\right] \, . \label{Q_t}
\ee

\section{Numerical results}
The field equations for the static radial Ansatz \re{MaxaxA}-\re{scalar_r} in constraint compliant parametrization consists of six second-order ODEs. The presence of $g(r)$, with the discontinuity in its boundary conditions at $|c_\infty|=|d_\infty|$, makes the integration of that system numerically challenging. In order to overcome such a difficulty, it is more convenient to use the Lagrange multiplier method, employing functions $R(r), S(r)$, and $T(r)$, instead, at the cost of introducing one more function, $\beta(r)$ (Lagrange multiplier), and the algebraic constraint \re{constraint_RST}. The system of seven ODEs plus an algebraic constraint, in eight functions, was solved by using the software package COLDAE, which employs a collocation
method for boundary-value differential-algebraic equations and a damped Newton method of quasi-linearization \cite{COLDAE}. In addition, a compactified radial coordinate $x=r/(1+r)$ was used.
Given a concrete theory (i.e., given $\mu$ and $\kappa$), there are three integer parameters ($n,m$, and $p$) and two real parameters ($c_\infty$ and $d_\infty$), which constitute our numerical set of parameters. Exploring the whole parameter space is an unattainable task, so we will concentrate on values that show the general behavior of excited solutions (satisfying \re{n_m_gt_2}).

\begin{figure}[ht!]
\begin{center}
 \includegraphics[height=.34\textwidth, angle =0 ]{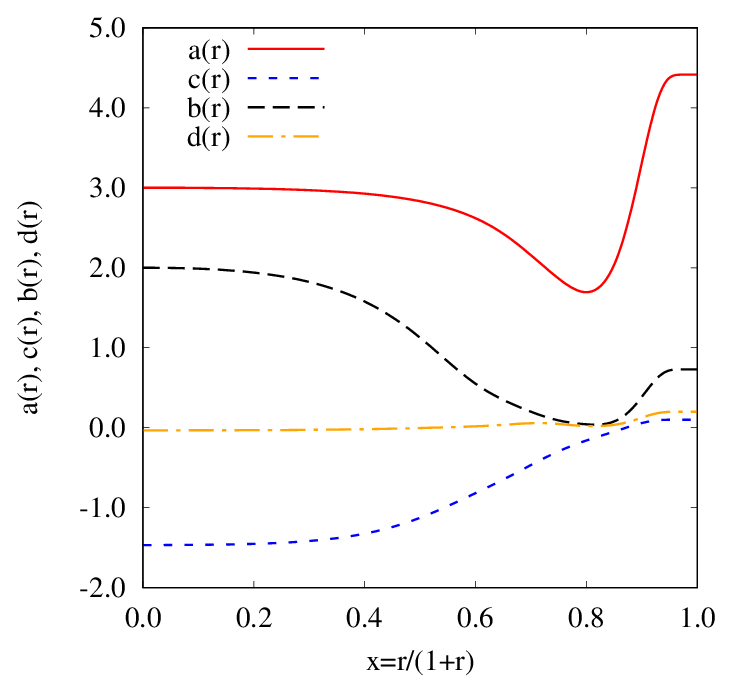} 
 \includegraphics[height=.34\textwidth, angle =0 ]{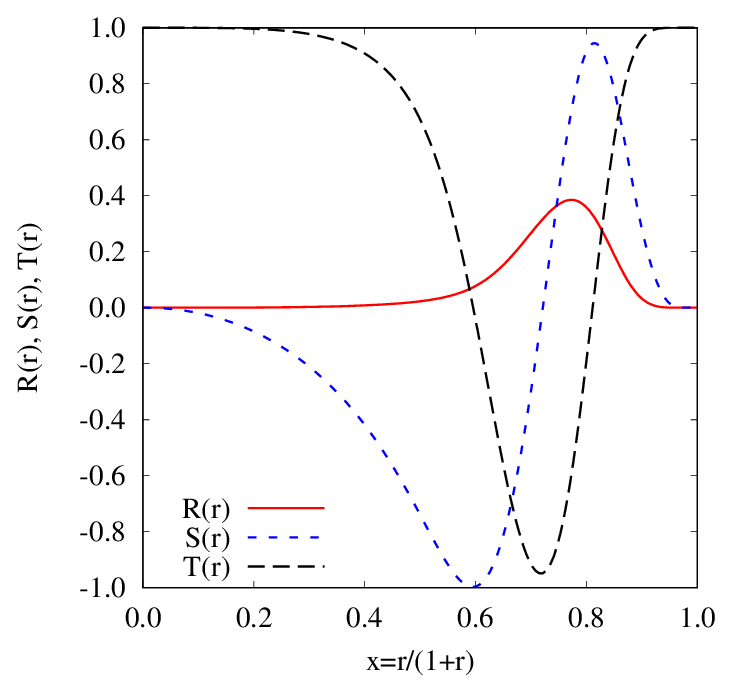}
\end{center}
\caption{  
The profiles of the functions of a typical solution are shown (for $\mu=0.1, \kappa=0.05, n=3, m=2, p=1, c_\infty=0.1, d_\infty=0.2$). 
}
\label{profile}
\end{figure}

\begin{figure}[ht!]
\begin{center}
 \includegraphics[height=.34\textwidth, angle =0 ]{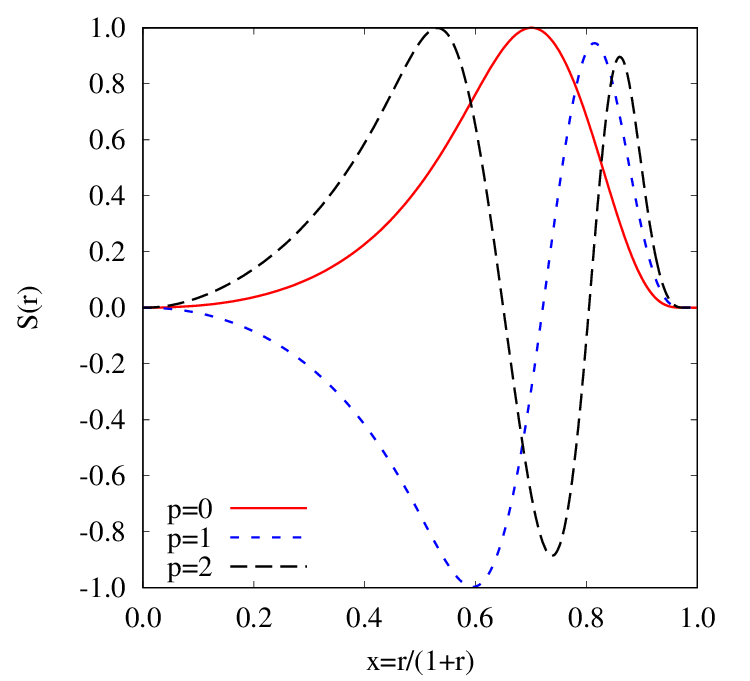}
 \includegraphics[height=.34\textwidth, angle =0 ]{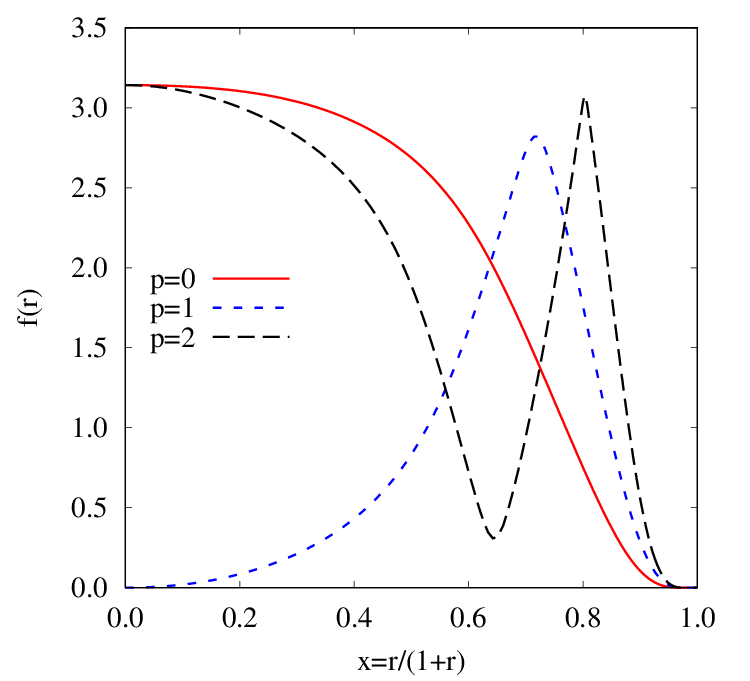} 
\end{center}
\caption{  
  Functions $S(r)$ (left) and $f(r)$ (right) for solutions with $\mu=0.1, \kappa=0.05, n=3, m=2, p=0,1,2, c_\infty=0.1, d_\infty=0.2$.
}
\label{nodes_p}
\end{figure}

The general profiles of the functions for a typical excited solution are presented in Fig.~\ref{profile}.  It is clear that the solutions interpolate between the boundary conditions at the origin  \re{BCs_A_0}-\re{BCs_RST_0} and at infinity  \re{BCs_A_infty}-\re{BCs_RST_infty}. The parameter $p$ can be interpreted as the number of nodes of $S(r)$ in $(0,\infty)$. In Fig.~\ref{nodes_p} (left) we exemplify that by considering three solutions with $p=0,1,2$, respectively. Notice that $p=0$ implies $g(r)=0$ (which gives rise to $R(r)=0$.) Then, these solutions with $p=0$ are fundamental embedded $O(3)$ solutions, since the symmetry group for the Skyrme scalar reduces to an effective $O(3)$ group. Excited solutions correspond to $p=1,2, \dots$. Fully $O(5)$ solutions have to be excited solutions, although there are excited  embedded $O(3)$ solutions for $p\neq 0$, though (see below). 

\begin{figure}[ht!]
\begin{center}
 \includegraphics[height=.34\textwidth, angle =0 ]{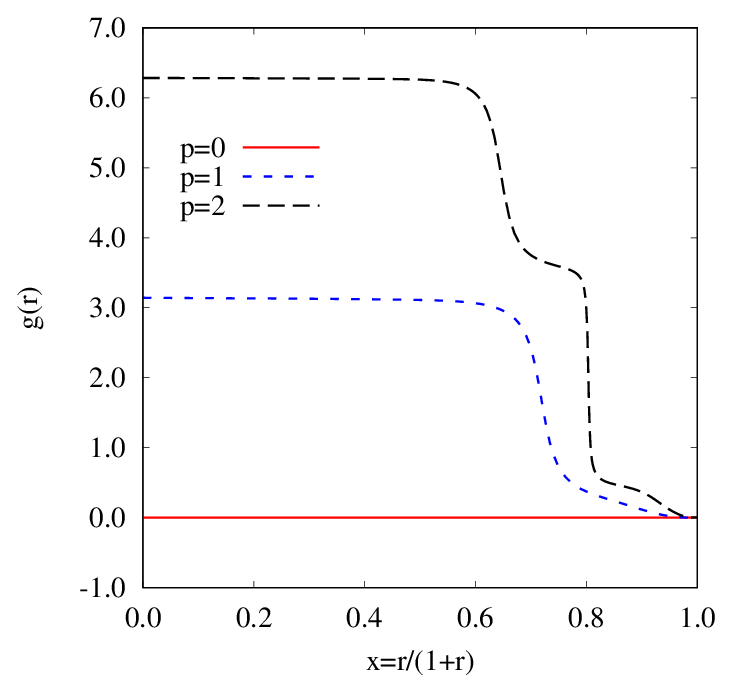}
 \includegraphics[height=.34\textwidth, angle =0 ]{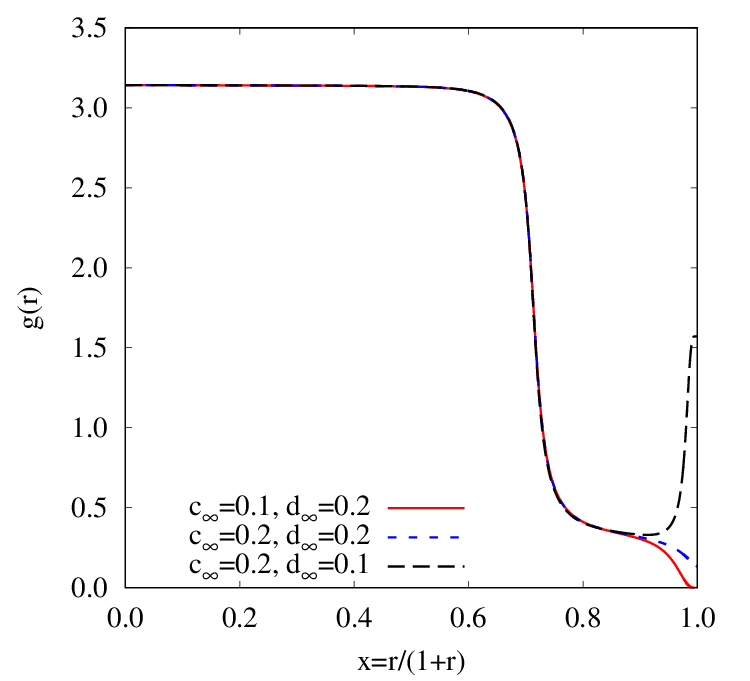} 
\end{center}
\caption{  
(left) Function $g(r)$ for solutions with $\mu=0.1, \kappa=0.05, n=3, m=2, p=0,1,2, c_\infty=0.1, d_\infty=0.2$.
  (right)  Function $g(r)$ for solutions with $\mu=0.1, \kappa=0.01, n=3, m=2, p=1, (c_\infty=0.1, d_\infty=0.2), (c_\infty=0.2, d_\infty=0.2)$, and $(c_\infty=0.2, d_\infty=0.1)$.
}
\label{function_g}
\end{figure}

In Fig.~\ref{nodes_p} (right) we give the corresponding profiles of $f(r)$ for the solutions presented in  Fig.~\ref{nodes_p} (left). It is seen that as $p$ increases the last peak of $f(r)$ approaches the value $\pi$, never being reached though. The corresponding profiles for function $g(r)$ are presented in Fig.~\ref{function_g} (left). Again, the parameter $p$ can also be interpreted as the number of ``steps'' the function $g(r)$ shows.

As mentioned previously, using $g(r)$ for numerics introduces additional difficulties due to the discontinuity that occurs at $|c_\infty|=|d_\infty|$ over the boundary conditions at infinity. That jump is shown in Fig.~\ref{function_g} (right). As $|c_\infty|$ and $|d_\infty|$ get closer, $g(r)$ becomes more and more steep at $x=1$, making necessary the use of a non-constraint compliant parametrization.

The full parameter space shows very intricate patterns depending on the values of $n$, $m$, and $p$, as well as of $\mu $ and $\kappa$. Here we will show the typical general behavior of the global quantities defined above. In order to do so, we will consider two types of curves: curves with constant $c_\infty$ and varying $d_\infty$ and curves with constant $d_\infty$ and varying $c_\infty$.

\begin{figure}[ht!]
\begin{center}
 \includegraphics[height=.35\textwidth, angle =0 ]{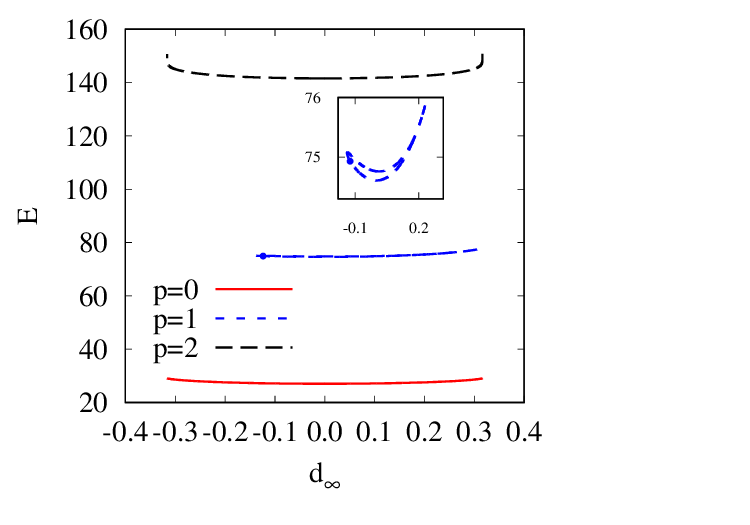}
 \includegraphics[height=.34\textwidth, angle =0 ]{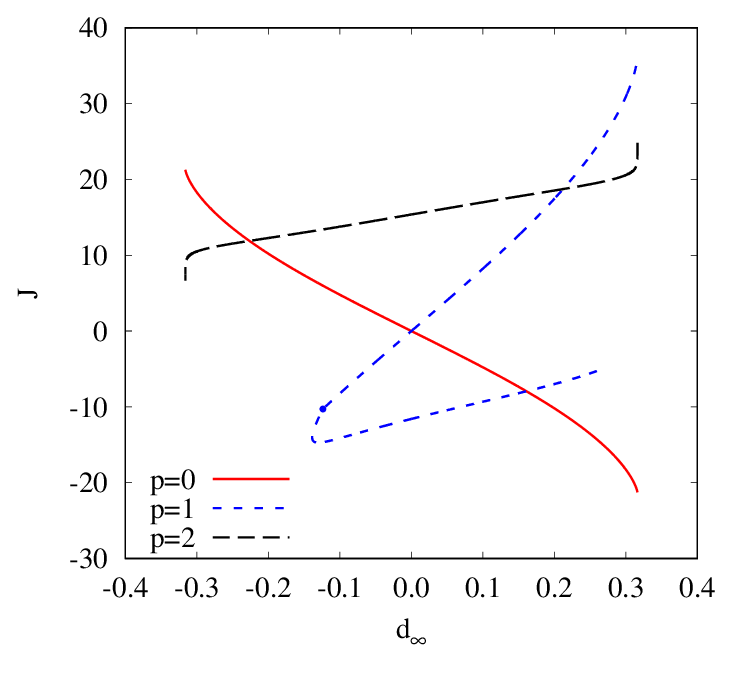} 
\end{center}
\caption{  
(left) Energy $E$ vs $d_\infty$ for solutions with $\mu=0.1, \kappa=0.05, n=3, m=2, p=0,1,2, c_\infty=0.1$. (right) Angular momentum $J$ vs $d_\infty$ for solutions with $\mu=0.1, \kappa=0.05, n=3, m=2, p=0,1,2, c_\infty=0.1$.
}
\label{E_J_d_infty}
\end{figure}

\begin{figure}[ht!]
\begin{center}
  \includegraphics[height=.34\textwidth, angle =0 ]{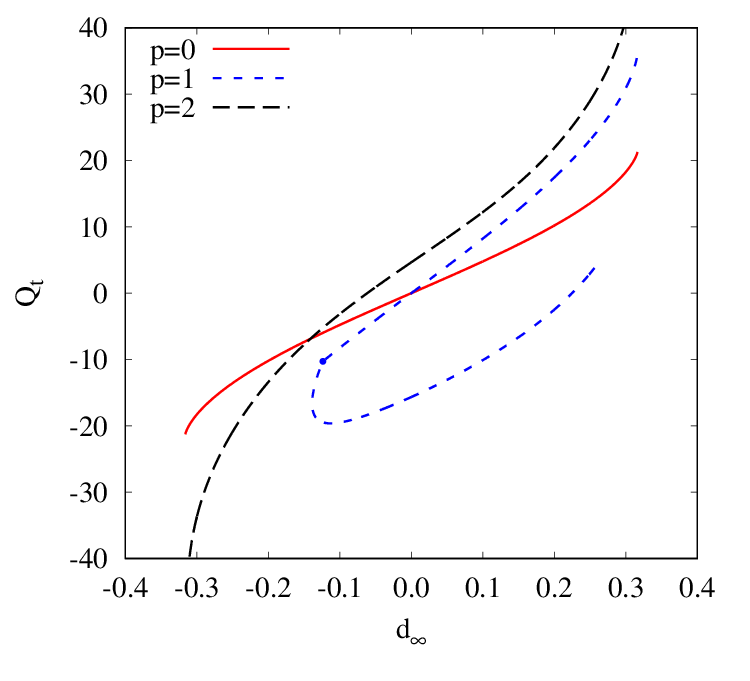}
   \includegraphics[height=.34\textwidth, angle =0 ]{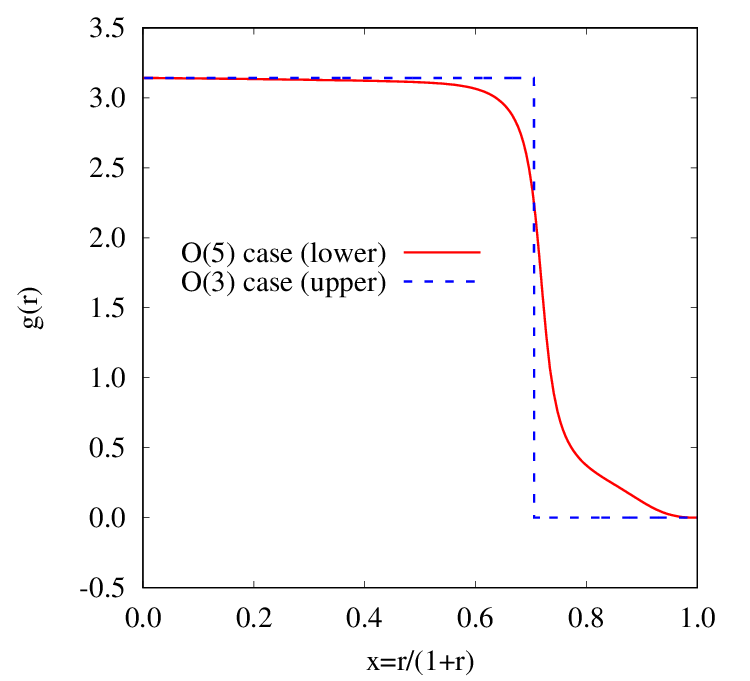}
\end{center}
\caption{  
(left) Total electric charge $Q_t$ vs $d_\infty$ for solutions with $\mu=0.1, \kappa=0.05, n=3, m=2, p=0,1,2, c_\infty=0.1$. (right) Function $g(r)$ for solutions with $\mu=0.1, \kappa=0.05, n=3, m=2, p=1, c_\infty=0.1, d_\infty=0.2$, both for fully $O(5)$ Skyrme scalar (corresponding to the lower branch in left figure) and for embedded $O(3)$ Skyrme scalar (corresponding to the upper branch in left figure).
}
\label{Q_t_d_infty}
\end{figure}

In Fig.~\ref{E_J_d_infty} we represent the energy $E$ and the angular momentum $J$ as a function of $d_\infty$ for solutions with $\mu=0.1, \kappa=0.05, n=3, m=2, p=0,1,2, c_\infty=0.1$. In both plots it is clearly seen the different behavior between solutions with odd $p$ ($p=1$ in the plots) and even $p$ ($p=0,2$ in the plots). While for even $p$ the behavior of $E$ and $J$ is the expected one, with solutions existing in the range set by the asymptotic behavior ($\mu \ge d_\infty^2$), for $p=1$ we observe two branches, corresponding to two different solutions for each given $d_\infty$. This same feature also occurs for the total electric charge $Q_t$, shown in Fig.~\ref{Q_t_d_infty} (left). In this figure, along the lower branch and up to the bullet on the upper branch, the solutions correspond to fully $O(5)$ Skyrme scalar field while solutions on the upper branch to the right of the bullet correspond to embedded $O(3)$ Skyrme scalar solutions \footnote{Notice that the position of the lower and upper branches are inverted for the energy $E$, the ``upper branch'' being below the ``lower branch'' in the plot for $E$.}. For these solutions $R(r)=0$. The embedded $O(3)$ solutions are excited, though, since $S(R)$ has one node, as they correspond to $p=1$. It is worth noticing that the $O(3)$ solutions along the upper branch cannot be computed in constraint compliant parametrization \re{cct}. The reason for that is that function $g(r)$ is not continuous, having a jump from $\pi$ to $0$ at a certain value of the radial coordinate. In  Fig.~\ref{Q_t_d_infty} (right) we plot $g(r)$ for solutions with $d_\infty=0.2$, both along the lower branch (fully $O(5)$ case) and the upper branch (embedded $O(3)$ case), revealing clearly the discontinuity of $g(r)$ for the $O(3)$ solutions.

\begin{figure}[ht!]
\begin{center}
 \includegraphics[height=.35\textwidth, angle =0 ]{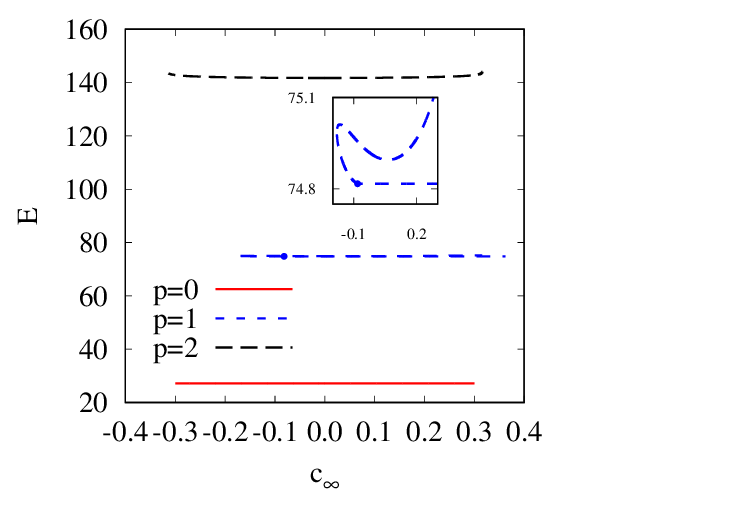}
 \includegraphics[height=.34\textwidth, angle =0 ]{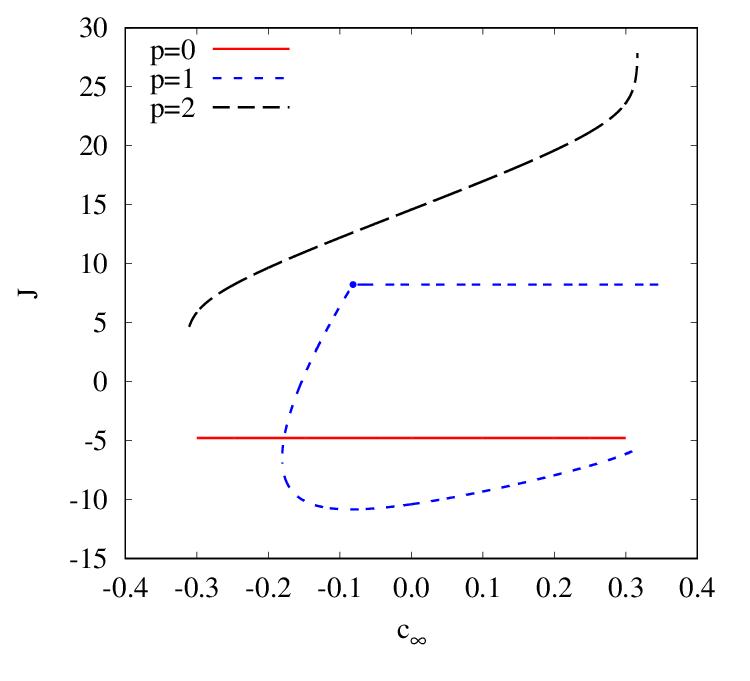} 
\end{center}
\caption{  
(left) Energy $E$ vs $c_\infty$ for solutions with $\mu=0.1, \kappa=0.05, n=3, m=2, p=0,1,2, d_\infty=0.1$. (right) Angular momentum $J$ vs $c_\infty$ for solutions with $\mu=0.1, \kappa=0.05, n=3, m=2, p=0,1,2, d_\infty=0.1$.
}
\label{E_J_c_infty}
\end{figure}

\begin{figure}[ht!]
\begin{center}
  \includegraphics[height=.34\textwidth, angle =0 ]{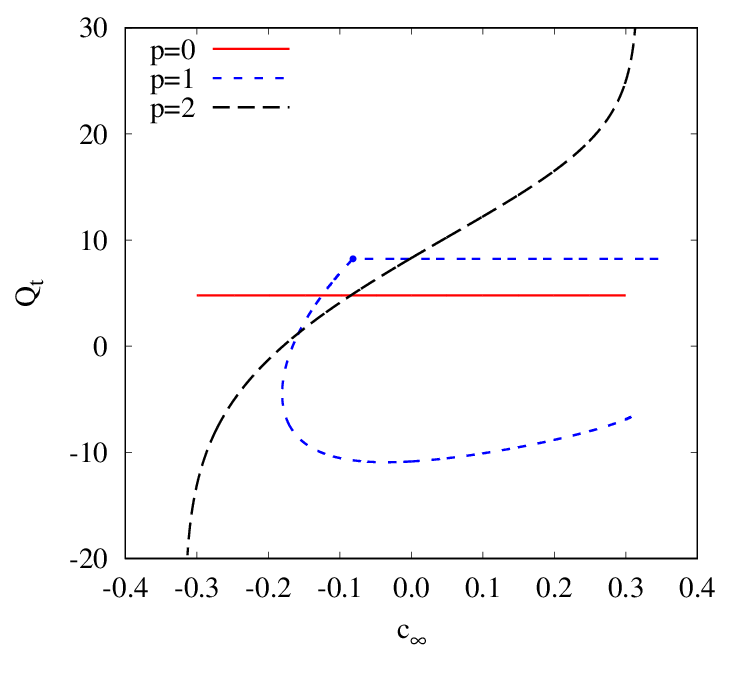}
  \includegraphics[height=.34\textwidth, angle =0 ]{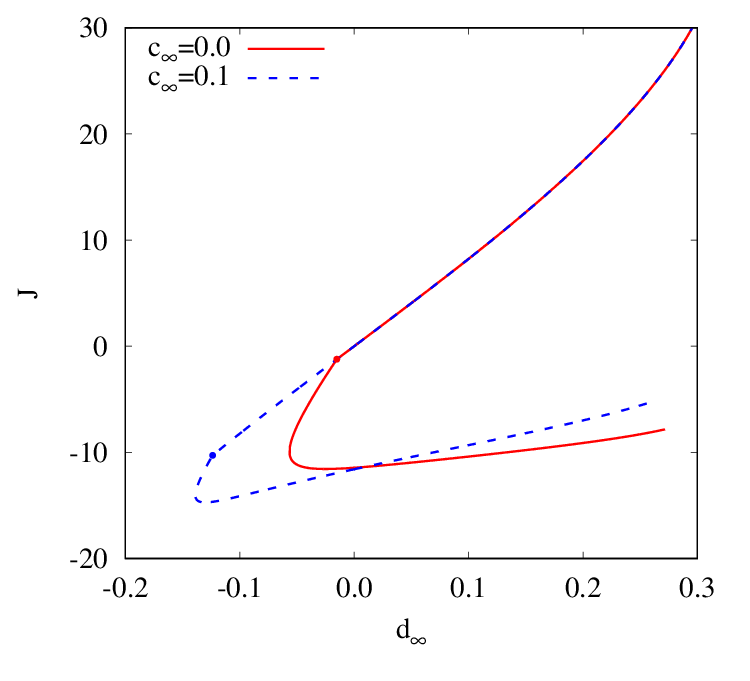}
\end{center}
\caption{  
  (left) Total electric charge $Q_t$ vs $c_\infty$ for solutions with $\mu=0.1, \kappa=0.05, n=3, m=2, p=0,1,2, d_\infty=0.1$. (right) Angular momentum $J$ vs $d_\infty$ for solutions with $\mu=0.1, \kappa=0.05, n=3, m=2, p=1, c_\infty=0.0, 0.1$. 
}
\label{Q_t_c_infty}
\end{figure}

In Figs.~\ref{E_J_c_infty} (left and right) and Fig.~\ref{Q_t_c_infty} (left), $d_\infty$ is set to $d_\infty=0.1$ and $c_\infty$ is varied, for solutions with $\mu=0.1, \kappa=0.05, n=3, m=2, p=0,1,2$. Again the different behavior between even and odd $p$ solutions appears. In this case, however, the $O(3)$ solutions that appear to the right of the bullet have constant global quantities. In fact, all the points along the horizontal lines represent just one solution. The reason for that is that $c_\infty$ becomes a pure gauge parameter for embedded $O(3)$ solutions (i.e., solutions with $p=0$ and $O(3)$ solutions along the upper branch). This is checked in Fig.~\ref{Q_t_c_infty} (right) where we represent the angular momentum $J$ as a function of $d_\infty$ for two values of $c_\infty$, namely, $c_\infty=0.0$ and $0.1$, for solutions with $\mu=0.1, \kappa=0.05, n=3, m=2, p=1$. We clearly see that both cases have a common part along the embedded $O(3)$ solutions (upper branches), reflecting the fact that $c_\infty$ is a gauge parameter for those embedded $O(3)$ solutions. Notice that these embedded $O(3)$ solutions are excited solutions ($p=1$), though.

Moreover, we should notice that using $p>0$ as the label for excited solutions is in order, since for given $\mu, \kappa, n, m, c_\infty$, and $d_\infty$, we have always found numerically that the energy $E$ increases with $p$, as seen in Fig.~\ref{E_J_d_infty} (left) and Fig.~\ref{E_J_c_infty} (left). Although we have not been able to explore the whole parameter space, this seems to be a general property for these solutions.

\begin{figure}[ht!]
\begin{center}
  \includegraphics[height=.34\textwidth, angle =0 ]{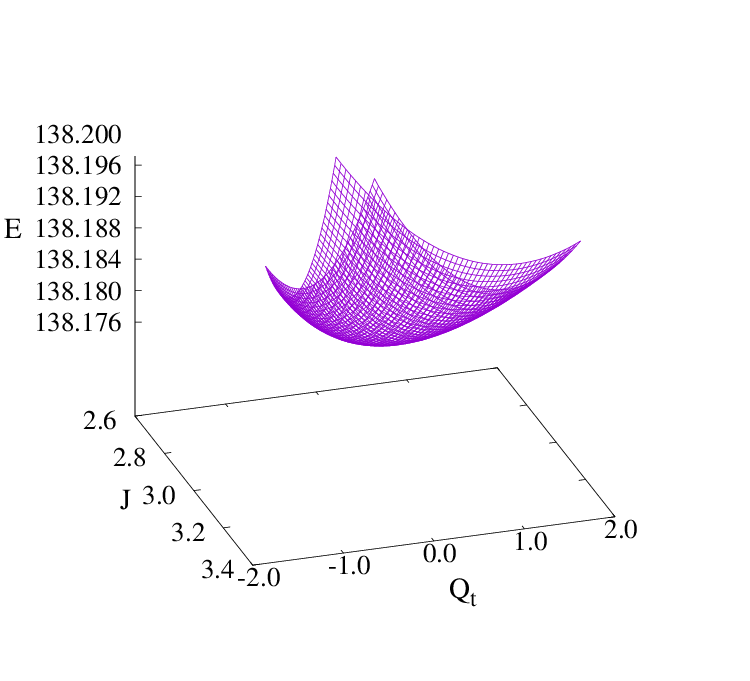}
  \includegraphics[height=.34\textwidth, angle =0 ]{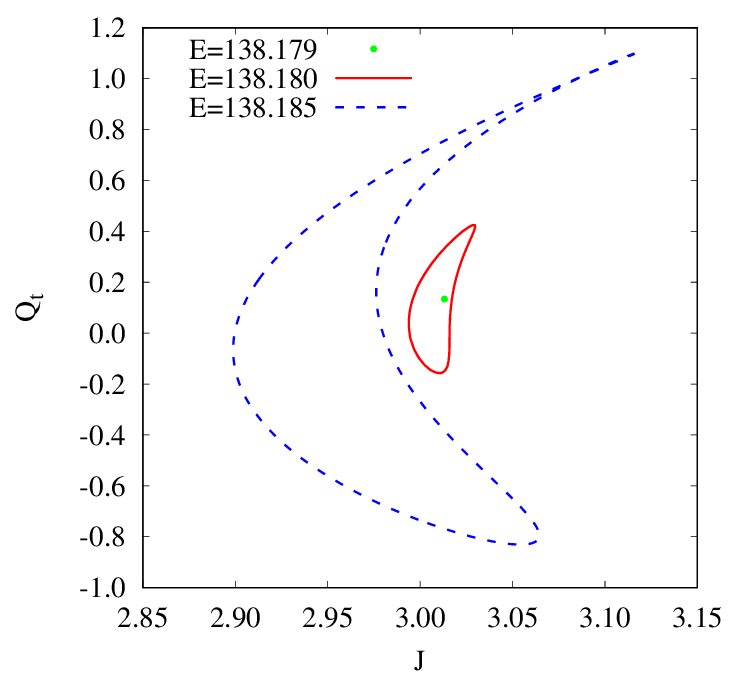}
\end{center}
\caption{  
  (left) Energy $E$ vs $(J, Q_t)$ for solutions with $\mu=0.1, \kappa=0.05, n=3, m=2, p=2$. (right) Contour curves for the left surface around the local minimum for $E=138.179, E=138.180$, and $E=138.185$. 
}
\label{E_J_Q_t}
\end{figure}

As a final remark, we would like to mention that for given $(n, m, p, \kappa, \mu)$, there is a solution for which the energy $E$ reaches a minimum. Although we have not analyzed all the possibilities, we have always found that a local minimum occurs at $(c_\infty,d_\infty)=(0,0)$. For $p=0$ that corresponds to the solution with $J=0, Q_t=0$, but for $p=1,2, \dots$, the minimum of $E$ corresponds to a charged rotating solution, i.e., with $J \neq 0$ and $Q_t \neq 0$. As an example of that, we show in Fig.~\ref{E_J_Q_t} (left) the dependence of the energy $E$ on the angular momentum $J$ and the total electric charge $Q_t$ in the vicinity of the minimum for the solutions with $\mu=0.1, \kappa=0.01, n=3, m=2, p=2$. The shape of the surface $E(J,Q_t)$ is quite involved, as it is a multivaluated function. In Fig.~\ref{E_J_Q_t} (right) we display some contour curves for the $3D$ surface of  Fig.~\ref{E_J_Q_t} (left). It is evident from both figures that the energy $E$ is not a monotonous function neither on the angular momentum $J$ nor on the electric charge $Q_t$, for $p>0$. 

\section{Conclusions}
In this paper we have addressed the construction of excited solutions in a SCS model in $2+1$ dimensions and the question whether that SCS term might modify or not the general pattern of energies in the model, by analyzing its effect on the excited solutions. Although excited solutions (with $p>0$) already exist in the ungauged Skyrme model in $2+1$ dimensions (provided that the quartic Skyrme kinetic term is present in the Lagrangian, \re{L}), here we have introduced an $SO(2) \times SO(2)$ Abelian gauge potential to be able to define a SCS term, whose influence we were aimed to study.

In Ref.~\cite{Navarro-Lerida:2023fsr} the effect of a SCS term on fundamental solutions of a slightly different model was considered, revealing that the slopes of $(E, Q_e)$ and $(E, J)$-curves may have both positive and negative signs, in contrast with the standard case of exclusively monotonically increasing positive slopes. However, only solutions with $g(r)=const.$ were considered there. Our objective here was to look for the existence of excited solutions first and then study their properties. Their construction was challenging, mainly because the use of a constraint compliant parametrization introduces a discontinuity in the asymptotic boundary condition of the function $g(r)$, and also because excited embedded $O(3)$ solutions in constraint compliant parametrization have a non-continuous $g(r)$. That was the reason why the type of solutions we are considering here could not be obtained in Ref.~\cite{Navarro-Lerida:2023fsr}.   

Using the Lagrange multiplier method, the excited solutions can be obtained, being characterized by an integer number $p$, which coincides with the number of nodes of the function $S(r)$ in $(0,\infty)$. Fundamental solutions correspond to $p=0$. In addition, the Ansatz considered here, \re{MaxaxA}-\re{constraint_RST}, includes two more integer numbers, namely, $n$ and $m$. Excited solutions exist only for $|n| \neq |m|$ and $|n|, |m| \ge 2$. There is an asymmetry between solutions with even $p$ and odd $p$, having the latter a more intricate behavior (see Figs.~\re{E_J_d_infty} and \re{Q_t_d_infty}). In fact, we have only found excited embedded $O(3)$ solution for odd $p$ (Figs. \re{E_J_d_infty}-\re{Q_t_c_infty}).

Finally, for given $n$, $m$, $\kappa$, and $\mu$, we observe, in general, that the energy increases with $p$, the lowest value corresponding to the fundamental solutions $p=0$ (see Fig.~\re{E_J_d_infty}). Although it is impossible to explore the whole parameter space of the theory, we believe this is a general property. In that sense, the SCS term does not seem to be enough to reverse the general order of the energy levels in a gauged Skyrme model in $2+1$ dimensions.

\section*{Acknowledgments}
F.N.-L. gratefully acknowledges support  from MICINN under project PID2021-125617NB-I00 ``QuasiMode".

\begin{small}

\end{small}
\medskip
\medskip

\end{document}